\documentstyle[aasms4]{article}
\def\ls{{_<\atop^{\sim}}}
\def\gs{{_>\atop^{\sim}}}
\def\pm{{_+\atop^{-}}}
\begin{document}

\title{Resonant Absorption in the AGN spectra emerging from photoionized gas: 
differences between steep and flat ionizing continua}

\author{ Fabrizio Nicastro$^{1,2,3}$, Fabrizio Fiore$^{1,2,4}$, 
and Giorgio Matt$^5$}

\affil {$^1$ Harvard-Smithsonian Center for Astrophysics \\
60 Garden st. Cambridge Ma. 02138 USA}

\affil {$^2$ Osservatorio Astronomico di Roma \\
via Osservatorio, Monteporzio-Catone (RM), I00040 Italy}

\affil {$^3$ Istituto di Astrofisica Spaziale - CNR \\
Via del Fosso del Cavaliere, Roma, I-00133 Italy}

\affil {$^4$ BeppoSAX Science Data Center\\
via Corcolle 19, Roma I00100 Italy}

\affil {$^5$ Dipartimento di Fisica, Universit\`a  degli Studi ``Roma Tre''\\ 
Via della Vasca Navale 84, Roma, I00146 Italy}

\author{\tt version: 18 November 1998}

\begin {abstract}

We present photoionization models accounting for both 
photoelectric and resonant absorption.
Resonance absorption lines from C, O, Ne, Mg, Si S and Fe between 
0.1 and 10 keV are treated. In particular we consider 
the complex of almost 60 strong Fe L absorption lines around 1 keV. 
We calculate profiles, intensities and equivalent widths of each line, 
considering both Doppler and natural broadening mechanisms. 
Doppler broadening includes a term accounting for turbulence of the 
gas along the line of sight. 

We computed spectra transmitted by gas illuminated
by drastically different ionizing continua and compared them
to spectra observed in flat X-ray spectrum, broad optical emission line
type 1 AGN, and steep X-ray spectrum, narrow optical emission line type 
1 AGN. We show that the $\sim 1$ keV absorption feature observed in 
moderate resolution X-ray spectra of several Narrow Line Seyfert 1 
galaxies can be explained by photoionization models, taking into account 
for resonance absorption, without requiring relativistic outflowing 
velocities of the gas, if the physical properties of these absorbers 
are close to those found in flat X-ray spectrum Seyfert 1 galaxies.

We finally present simulations of the spectra emerging from gas 
illuminated by both steep and flat ionizing continua, as seen by the 
AXAF High energy Transmission Gratings and the baseline Constellation-X 
calorimeter. We discuss briefly the relevant physics which can be 
investigated with these instruments. 

\end{abstract}

\section{Introduction}

The ROSAT (Tr\"umper 1983)  PSPC (Pfefferman et al. 1987) 
has found a large spread in the spectral indices of type 1 AGN: 
$0.5<\alpha_E<3$ ($f(E)=E^{-\alpha_E}$; Laor et al 1997, Walter \& 
Fink, 1993, Fiore et al. 1994).  This remarkably broad soft X-ray
spectral index distribution appears not to be at random, but rather it
is strongly correlated with other important optical and X-ray
properties of these objects. Objects with narrower H$\beta$ ($1000<$
FWHM$<2000$ km s$^{-1}$) and smaller [OIII]/H$\beta$ ($<1$) ratio are 
systematically steeper in soft X-rays ($\alpha_E > 1.5$) than AGN with 
broad (FWHM$> 2000$ km s$^{-1}$) permitted lines and greater [OIII]/H
$\beta$ ratio ($1<$[OIII]/H$\beta<3$, $\alpha_{PSPC} <1.5$). 
In particular most of the so called Narrow Line Seyfert 1
Galaxies (NLSy1) show a steep 0.2-2 keV continuum (Boller et al 1996).
In the 2--10 keV energy range narrow optical emission line quasars have 
still a somewhat steeper spectrum compared to broad optical emission 
line quasars. 
The spread of the 2-10 keV spectral index is however smaller than 
that of the 0.1-2 keV, PSPC one (Brandt et al., 1997).  
It has been suggested that these properties are correlated with the 
accretion rate (Pounds et al 1995, Laor et al 1997, Fiore et al 1998a,b).

Ionized gas, modifying the AGN X-ray spectrum, has been
detected in about half of the bright, X-ray flat, broad optical emission 
line type 1 AGN studied by ASCA (Reynolds, 1997, George, 1998). 
The main features imprinted by the ionized gas on the X-ray spectrum are 
the OVII and OVIII K absorption edges at 0.74 and 0.87 keV. 
{\bf As already pointed out by George et al. (1998), the similarity of the 
parameters of these features among different sources in the both the 
Reynolds (1997) and the George et al. (1998) samples implies that 'warm' 
absorbers in AGN have all roughly similar ionization structure, despite of 
the broad range of ionizing luminosity (i.e. about 3 orders of magnitude, 
from $\sim10^{41}$ to $10^{44}$ erg s$^{-1}$).}
In terms of the ionization parameter U, the presence in the gas of 
He-like and H-like ions of C, O and Ne as dominant ionic species, implies 
a narrow range of LogU=0.5-1.5 (Nicastro et al, 1998a).  
This is consistent with the location of the 'warm' gas at radii scaling 
with the root square of the luminosity. 
The column densities of the Reynolds sample are distributed in the range 
log$N_H$=21-23, but the standard deviation of the distribution is rather 
small, $\sigma_{{\rm log}N_H}=0.3$, again indicating a relative uniformity 
in the physical properties of this AGN component. 
Several type 1 AGN showing ionized absorption features in X-ray also show 
high ionization UV absorption lines (namely OVI, NV, CIV, CIII]), 
over-imposed to the broad emission lines and systematically blue-shifted 
with respect to them (Mathur et al., 1994, 1995, 1998). {\bf The widths 
of these absorption lines are tipically much broader than the associated 
thermal broadening, suggesting turbulence velocities of the absorber.}  
The UV and the X-ray absorbers have ionization states compatible with each 
other, which supports the identification of these two components (Mathur 
et al., 1994, 1995, 1997, 1998). 
{\bf In one of these cases (the Seyfert 1 galaxy NGC~3516) the total column 
density implied by some strong UV absorption lines is larger than that 
required by the X-ray absorption features (Kriss et al., 1996a,b). 
However it is well possible that more than one kinematic components 
contribute to the total intensity and equivalent width of the UV lines 
and that only one of those is physically associated with the X-ray absorber 
(Mathur et al., 1997).} 
The UV absorption lines may be produced in a turbulent outflowing wind 
intercepting the line of sight (possibly driven out by radiative 
acceleration). 
The outflowing velocities of the UV 'warm' gas span the range $v_{out}\sim 
100 - 2000$ km s$^{-1}$ (Mathur et al. 1994, 1995, 1997), and a similar 
range has been estimated for the dispersion velocities $\sigma_{v}$ along the
line of sight.

\bigskip
On the other hand, deep absorption features at the energies of the K
edges of high ionized oxygen have never been observed in the 0.1-2 keV
spectra of soft X-ray steep spectrum, narrow optical emission line 
type 1 AGN.  
Detections of deep oxygen edges have been reported for two sources previously
classified as NLSy1 (IRAS~17020+4544, Leighly et al., 1997; Mrk~766,
Matt et al., 1998, RE~J2248-511, Puchnarewicz et al., 1995).  
During the ASCA (0.5-10 keV) and BeppoSAX (0.1-10 
keV) observations the two sources have a rather flat X-ray continuum 
($\alpha_E\sim1$) and therefore cannot be considered X-ray steep AGNs. 
Large variations of the X-ray spectral shape have already been 
observed in the past (e.g. 1H~0419-577, NGC~4051, RE~J2248-511, 
Guainazzi et al 1998a,b, Turner et al 1998, Puchnarewicz et al., 1995). 
{\bf It is well possible that a population of (perhaps low mass) AGN does 
exist, switching between soft, high states and hard, low states 
(depending possibly on the accretion rate), and therefore classified in a 
way or in the other depending on the observed state (in analogy to 
what seen in Galactic Black Hole Candidates).} 
 
Despite the lack of oxygen edges, steep soft X-ray spectrum, narrow 
optical emission line, type 1 AGN are usually far from being featureless 
(Brandt, 1995; Hayashida, 1997).  
ASCA has found large residuals between 1 and 2 keV, after subtracting 
the appropriate continuum, in at least 4 cases (Leighly et al., 1997, 
Fiore et al., 1998a). 
The interpretation of these features is not unique, given 
the relatively modest spectral resolution of the ASCA SIS and the 
complexity of the continua.
Leighly et al. (1997) interprets the features found in the three NLSy1
studied in terms of OVII and OVIII resonance absorption lines blushifted 
by $\sim 0.5$ c.  Fiore et al. (1998a) interpret the feature found in 
PG1244+026 in terms of either an emission line at 0.92 keV or Fe XVIII 
and Fe XVII resonant absorption blueshifted by $\sim 0.3$ c.

Photoionization models used to fit and interpret AGN spectra 
usually deal with photoelectric absorption only. Matt (1994) and 
Krolik \& Kriss (1995) were the first to point out the importance
of X--ray resonant absorption lines in ionized absorbers. 
Here we calculate in detail  photoionization models, including all
the relevant resonant absorption lines, and study the dependence of the 
transmitted spectrum on the exact shape of the ionizing continuum. 
We find that the phenomenologies observed in both broad optical emission 
lines type 1 AGN and NLSy1 galaxies can be explained in terms of their 
quite different ionizing continua, provided that, for a given luminosity, 
the radial distance of the gas clouds from the ionizing source (in units of 
gravitational radii) and the gas density are similar in both classes 
of objetcs. 

Netzer (1993, 1996) discussed the relevance of recombination and 
fluorescent emission lines, provided that the covering factor of the
ionized matter is significant, while Krolik \& Kriss (1995) and 
Matt, Brandt \& Fabian (1996) pointed out that resonant scattering lines
can be even more intense. Here, for the sake of simplicity, we do not 
consider any line emission, and therefore our model is strictly
valid only if the covering factor is negligible (see next section). 

In section 2 we present our photoelectric+resonant absorption
model, and discuss the relevant physics.  Section 3 compares the
transmitted spectra computed with our models for two different
ionizing continua with the spectra of a broad optical emission line 
type 1 AGN and a NLSy1 galaxy actually observed by ASCA ($\Delta E = 
100$ eV, effective area $\sim 500$ cm$^{2}$).  
Section 4 presents a set of simulations with instruments of much higher 
spectral resolution (AXAF High Energy Transmission Gratings, HETG, 
$\Delta E = 1.5$ eV, collecting area $\sim 100$ cm$^2$, 
``AXAF Proposer's Guide'', vs 1.0, 1997 
\footnote{
http://asc.harvard.edu/USG/docs/pg\_c\_ps.html
}
; and Constellation-X baseline Calorimeter, $\Delta E = 3$ eV, 
collecting area $\sim 10,000$ cm$^2$, ``The High Throughput X-ray 
Spectroscopy (HTXS) Mission'', 1997
\footnote{
http://constellation.gsfc.nasa.gov/documentation.html.
}
). We finally present our conclusions in Section 5.

\section{Models} 

We calculated the curves of growth of all the 268 permitted resonant 
absorption lines of C, O, Ne, Mg, Si, S and Fe, in the 0.1--10 keV energy 
range, with oscillator strengths greater than 0.05. 
{ \bf We used atomic data from the complete list of permitted LS-type 
(LS coupling transitions) resonant lines of all ionization states of 18 
elements tabulated in Verner et al. (1996). 
Intercombination and forbidden lines are not included in that compilation.}
We carried out the calculations considering the exact Voigt profile for
the lines, i.e. considering both Doppler and natural broadening. 
As suggested by the width of the absorption lines detected in the UV spectra 
of the UV/X-ray ionized absorbers (Mathur et al. 1994, 1995, 1997, 1998), 
in addition to the thermal Doppler broadening we consider here a 
term accounting for gas turbulence, and refer to this term as the 
``turbulence velocity'' $\sigma_v$ along the line of sight. 

We calculated spectra emerging from photoionized gas for a number of 
values of N$_H$, U, and $\sigma_v$. Relative ionic abundances and edge-like 
absorption features for given U and N$_H$ and for a particular 
shape of the nuclear ionizing continuum are those calculated by CLOUDY 
(Ferland, 1997, vs 90.04), while the intensity and profile of the 
resonance absorption lines are consistently computed here for a given 
value of $\sigma_v$. 
The result is an emerging spectrum including both absorption edges 
and lines. 

We considered only line absorption, not emission. {\bf We used CLOUDY 
(vs 90.04, Ferland, 1997) to build total (transmitted + emitted) spectra 
and verified that, for the range of total column densities and ionization 
parameters considered here, this assumption corresponds (for a spherical 
configuration) to covering factors $f_c < 0.1$ (in good agreement 
with the Netzer, 1993, results: see his Fig. 3). 
In an ideal optically thin case, in which (a) the ionized gas isotropically 
and uniformely covers the central source ($f_c = 1$), (b) any systematic 
(inflowing or outflowing) motion of the gas is absent, and (c) 
any line destruction mechanism (autoionization, Compton scattering, 
photoabsorption) is inefficient, then scattering of the resonant emission line 
photons in the line of sight would exactly compensate for the line absorption 
along the line of sight, and, according to Netzer (1996), only emission 
features (mainly recombination lines) would be present in the total spectra. 
Any deviation from the above ideal condition would produce net absorption 
lines, which can therefore diagnose the physical and geometrical conditions 
of the absorbing material.}
Presence of outflows instead would imply that both emission and absorption 
lines are present in the emergent spectra, with the absorption lines 
blueshifted with respect to the emission ones. In this paper, however, for 
the sake of simplicity we assume that a cloud of gas with negligible 
covering factor {\bf ($f_c < 0.1$)} is completely obscuring the line of sight, 
and therefore that the contribution from emission lines to the overall 
spectrum is negligible. 
In comparing our model with real spectra it must be taken in 
mind that absorption lines may be partly compensated by line scattering 
{\bf and recombination lines} for significant covering factors, or that 
emission lines may appear redwards (bluewards) of the absorption lines if 
the matter is outflowing (inflowing). 

In the next paragraph we present the details of the computation, and show 
some examples of both growth curves and emerging spectra.

\subsection{Computation} 

The equivalent width of a resonant absorption line is: 
\begin{equation} \label{ew}
\mbox{EW} = \int_0^{+\infty} \left[ 1 - e^{-\tau_{\nu}} \right] d\nu, 
\end{equation}
where $\tau_{\nu}$ is the dimensionless specific optical depth of the 
considered transition: 
\begin{equation} \label{taunu}
\tau_{\nu} = \int_0^L ds \alpha_{\nu} = n_l L {{\pi e^2} \over {m c}} 
f_{lu} \Phi(\nu). 
\end{equation}
In the above equation $\alpha_{\nu}$ is the absorption coefficient at 
the frequency $\nu$ in cm$^{-1}$, $n_l$ is the number density of ions of 
the given element which populate the {\bf l}ower level in the unit volume (in 
cm$^{-3}$), L is the linear dimension of the cloud of gas along the line of 
sight (in cm), $f_{lu}$ is the oscillator strength of the electron 
transition from the {\bf l}ower to the {\bf u}pper level, m is the electron 
mass in g, and $\Phi(\nu)$ is the normalized Voigt profile.
With these units the equivalent width in equation (\ref{ew}) is in Hz. 
The Voigt profile can be written as (e.g. Rybicky \& Lightman 1979):
\begin{equation} \label{voigtpro}
\Phi(\nu) = {1 \over {\sqrt{\pi} \Delta\nu_D}} H(a,u),
\end{equation}
where $H(a,u)$ is the Voigt function: 
\begin{equation} \label{voigtfun}
H(a,u) = {a \over \pi} \int_{-\infty}^{+\infty} {{e^{-y^2} dy} \over 
{a^2 + (u-y)^2}}. 
\end{equation}
In equation (\ref{voigtfun}) $a$ is essentially the ratio between the 
natural and the Doppler line widths $\Gamma$ and $\Delta\nu_D$, and 
$u = (\nu - \nu_0) / \Delta\nu_D$. 
For small values of $a$ the center of the line is dominated by the 
Doppler profile, while the wings are dominated by the lorentzian profile. 
As is well known, this influences the growth curves making them increasing 
with N$_H$ even over the core saturation. 

\noindent
We consider two different contributions to the Doppler profile: (a) the 
thermal motion of the ions, and (b) the dispersion velocity $\sigma_v$ of 
the gas along the line of sight: 
\begin{equation} \label{dopplerwidth}
\Delta\nu_D = {\nu_0 \over c} \left( {{2kT} \over M} + \sigma_v^2 
\right)^{0.5}. 
\end{equation}
Let us estimate the relative contribution of the two terms in 
equation (\ref{dopplerwidth}): 
\begin{equation} \label{thermturb}
{{v^{therm}} \over {\sigma_v}} = {{\sqrt[2]{2 k T}} \over 
{\sqrt[2]{m} \sigma_v}} \simeq 1.3 {{(T)_6^{0.5} A^{-0.5}} \over 
{(\sigma_v)_{100}}}.  
\end{equation}
In equation (\ref{thermturb}) $(T)_6$ is the equilibrium temperature of 
the gas in units of $10^6$ K, $(\sigma_v)_{100}$ is the dispersion 
velocity in units of 100 km s$^{-1}$, and A is the atomic weight of the 
considered element. 
For a highly ionized gas in photoionization equilibrium, the temperature 
spans the range $(T)_6 \sim 0.1-1$. In a such a gas the elements lighter 
than O-Ne are usually fully stripped (the threshold element over which 
bound ions are still present depending strongly on the photoionizing 
continuum shape), and so the residual opacity from both bound--free and 
bound-bound transitions is mainly due to the heavy elements (from O to Fe). 
Assuming therefore $(T)_6=0.5$, $(\sigma_v)_{100} = 1$, and A=16 (the 
atomic weight of the lightest non-fully stripped element), we get, from 
equation (\ref{thermturb}), $v^{therm}/\sigma_v\simeq 0.2$. 
For the iron ions (A=56) we would find $v^{therm}/\sigma_v\simeq 0.1$. 
Hence, for dispersion velocities of $\sim 100$ km s$^{-1}$, the contribution 
of the thermal motion of the ions to the broadening of the absorption 
line is only $\sim 10-20\%$ of the total, and it decreases linearly with 
increasing $\sigma_v$. Therefore, the second term in equation 
(\ref{dopplerwidth}) is the dominant one for $(\sigma_v)_{100} \gs 1$.

\subsection{Growth Curves:}

Fig 1 and 2 show the growth curves of the K$\alpha$ and K$\beta$ lines of 
H-like and He-like C, O, Ne and Fe, for (a) $\sigma_v = 0$ (Doppler 
broadening entirely due to the thermal motion of the ions), and (b) $\sigma_v 
= 1000$ km s$^{-1}$ respectively. The equivalent widths of the two 
K$\alpha$ lines of H-like ions have been summed to give a single equivalent 
width for the doublet. 
%
%
%
%
The ionization state of the absorbing gas is that of a typical flat X-ray 
spectrum, broad optical emission line type 1 AGN 'warm absorber' (LogU=1) 
for C, O, and Ne, while a much higher ionization case (LogU=3) is adopted 
for the FeXXV and FeXXVI lines. 
In both cases the ionizing continuum  used is the 'AGN Spectral Energy 
Distribution' (SED), tabulated in CLOUDY (Ferland, 1997, vs. 90.04). 
However, we stress that the growth curves depend only on the relative 
abundance of the ion under investigations, and not on the particular 
combination of U and shape of the ionizing continuum producing this 
abundance. 

\noindent
At low column densities ($\ls 10^{21-23}$ cm$^{2}$: the exact value 
depending on the considered line and dispersion velocity), the equivalent 
width of each line increases linearly with N$_H$. 
For higher values of N$_H$ the optical depth of the line core saturates and 
the lorentzian profile begins to drive the further slower growth of the 
equivalent width with increasing columns. 
The linear branch of the growth curves (the low column density limit) is 
described by equation (\ref{ew}) in the limit $\tau_\nu << 1$: 
\begin{equation} \label{optthin}
\mbox{EW} \simeq \int_0^{+\infty} 
d\nu \tau_{\nu} = n_l L {{\pi e^2} \over {m c}} f_{lu} = \tau_{tot}.
\end{equation}
Let us define $A_X$, $\xi_{X^i}$ and N$_H$ as the abundance of the element 
X relatively to the H abundance, the relative abundance of the ion $i$ of 
the element X, and the equivalent hydrogen column density along the line 
of sight respectively. With these positions and introducing an appropriate 
numerical format we have: 

\begin{equation} \label{ewoptthin}
\mbox{EW} \simeq 2.2 (\xi_{X^i})_{0.4} (A_X)_{-5} (\hbox{N}_H)_{22} 
(f_{lu})_{0.5} \mbox{ eV}. 
\end{equation}

\noindent
When the core of the line becomes opaque to the radiation, equation 
(\ref{ewoptthin}) is no longer valid, and the computation of the 
equivalent width must be carried out using equation (\ref{ew}) with 
the exact shape of the profile function. 
A comparison between the curves in Fig. 1 and those in Fig. 2 allows 
to evaluate the effects of the line profile upon the resulting equivalent 
width. When the Doppler broadening is dominated by the thermal motion of 
the ions ($\sigma_v << 100$ km s$^{-1}$, Fig. 1), the dynamical range of 
variation of the equivalent width of each line with increasing N$_H$ is much 
smaller than in the case in which the Doppler broadening is dominated by a 
high dispersion velocity of the gas along the line of sight ($\sigma_v \gs 
100$ km s$^{-1}$). 
This is because when the line is strongly broadened the core of the line 
needs larger column of gas to become optically thick, and so the linear 
branch of the growth curves spans over a larger range of N$_H$ before 
saturating. 
From equations (\ref{taunu}), (\ref{voigtpro}) and (\ref{dopplerwidth}), 
using the same numerical factor as in equation (\ref{ewoptthin}), and 
setting $\beta_{\sigma_v} = \sigma_v/c$, we get: 
\begin{equation} \label{tau0}
\tau_0 \simeq 2.9 (f_{lu})_{0.5} (\xi_{X^i})_{0.4} (A_X)_{-5} A^{0.5}  
(\mbox{N}_H)_{22} \left[ (T)_6 + A (2.3\times10^3 \beta_{\sigma_v})^2 
\right]^{-0.5} E_0^{-1}.
\end{equation}
In the above equation the two terms within the square brackets describe the 
contribution to the core optical depth by the two considered Doppler 
mechanisms. From equation (\ref{thermturb}) we know that dispersion 
velocities of the gas along the line of sight greater than $\sim 100$ km 
s$^{-1}$ make the turbulence broadening mechanism very efficient compared 
to the thermal one, and so reduce the core optical depth of each line. 
Equation (\ref{tau0}) allows then to define and evaluate the core optical 
depth in the two extreme cases: $\sigma_v=0$ (thermal motion only: 
$\tau_0^{therm}$) and $\sigma_v \gg \sqrt{kT/M}$. We get: 
\begin{equation} \label{tau0therm}
\tau_0^{therm} \simeq 2.9 (f_{lu})_{0.5} (\xi_{X^i})_{0.4} (A_X)_{-5} A^{0.5}  
(\mbox{N}_H)_{22} (T)_6^{-0.5} E_0^{-1}, 
\end{equation}
\begin{equation} \label{tau0turb}
\tau_0^{turb} \simeq 1.24 (f_{lu})_{0.5} (\xi_{X^i})_{0.4} (A_X)_{-5}   
(\mbox{N}_H)_{22} (\beta_{\sigma_v})_{-3}^{-1} E_0^{-1}.
\end{equation}
The ratio between these two quantities does not depend on the particular 
considered transition, the distribution of the ionic abundances in the gas, 
and its column density, and is therefore a good estimator of the 
relative contribution of the two Doppler broadening mechanisms to the 
core optical depth: 
\begin{equation} \label{tau0thermtau0turb}
{{\tau_0^{therm}} \over {\tau_0^{turb}}} = 2.3 A^{0.5} (T)_6^{-0.5} 
(\beta_{\sigma_v})_{-3}. 
\end{equation}
Equation (\ref{tau0thermtau0turb}) tells us that a given column of 
non-turbulent highly ionized gas in photoionization equilibrium can 
be more than a factor 10 thicker to the resonant absorption 
process (at the energy of the transition) than an identical column of 
gas undergoing strong turbulence ($\sigma_v \gs 300$ km s$^{-1}$, i.e. 
$(\beta_{\sigma_v})_{-3} \gs 1$). 

\subsection{Emerging X-ray Spectra:}

The relative importance of the spectral features in spectra emerging 
from photoionized gas obscuring the line of sight strongly depends on 
the Spectral Energy Distribution of the ionizing continuum in 
the UV to X-ray energy range. 
We therefore treat separately the two extreme observed cases: (a) 
flat X-ray spectrum, broad optical emission line type 1 AGN continua, 
and (b) steep X-ray spectrum, narrow optical emission line type 1 AGN 
continua. 
{\bf If a peculiar class of AGN does exists, with intrinsic UV to X-ray 
continuum switching between these two extreme observed cases 
(as suggested by the behaviour observed in 1H~0419-577, NGC~4051, 
RE~J2248-511: Guainazzi et al. 1998a,b, Turner et al 1998, Puchnarewicz 
et al., 1995) then sources belonging to this class would show emerging 
spectra of the types (a) and (b) alternately.} 

\subsubsection{Flat X-ray Ionizing Continuum}

Typical O-UV to X-ray continua of broad line type 1 AGN (i.e. $\alpha_E 
\sim 1$, $\alpha_{OX} \sim 1.3$) transmitted by photoionized gas 
show deep OVII-OVIII edges, as those are, usually, the most abundant 
oxygen ions. 
In these conditions, when the Oxygen is mainly distributed in highly ionized 
states, Carbon is almost fully stripped and Neon is mostly distributed 
between NeIX-XI. 
Conversely high ionization Iron ions are not much populated, the main ions 
being FeXIV-XX (Nicastro et al., 1998a). 
In Fig. 3, we show 0.1--5 keV spectra emerging from an outflowing wind 
photoionized by a typical flat X-ray spectrum, broad optical emission 
line type 1 AGN continuum (Ferland, 1997), for three different values of 
LogU$ = 0.5, 1, 1.5$. 
We used an outflowing velocity of $v_{out} = 1000$ km s$^{-1}$, which is 
typical for the UV outflowing component in 'warm absorbers' (Mathur et al., 
1995), and which corresponds to  z $= -0.003$. For the dispersion velocity, 
we adopted a value $\sigma_v = 500$, km s$^{-1}$. 
The column of photoionized gas along the line of sight is fixed at N$_H = 6 
\times 10^{21}$ cm$^{-2}$, which is the mean value in the Reynolds (1997) 
sample. 
We also added absorption from cold gas of column N$_H^{Cold}=3\times10^{20}$ 
cm$^{-2}$ to simulate a typical Galactic absorption along the line of sight. 
In each panel of Fig. 3 we plot the transmitted spectra corresponding to a 
given value of LogU. 

%
%

Moving from LogU=0.5 to LogU=1.5 (i.e. from low to high ionization, 
from the upper to the lower panel of Fig. 3) the OVII K edge opacity 
decreases, due to the decreasing of the OVII/OVIII relative abundances ratio. 
Conversely the optical depth of the OVIII K edges increases until the 
OVII $\to$ OVIII photoionization rate becomes lower than the OVIII $\to$ 
OIX one. 
In the lower panel of Fig. 3 (LogU=1.5) only the OVIII K edge is visible, 
while the gas is almost completely transparent at the OVII K edge energy. 
In the lower ionization case (LogU=0.5: upper panel of Fig. 3), a deep 
OVII K edge is instead present, along with the 0.49 keV K edge of CVI. 
NeIX K edge is marginally visible because of the low column density adopted. 

Strong OVII K$\alpha$ and K$\beta$ and OVIII $K\alpha$ resonance absorption 
lines at 0.574, 0.653, and 0.665 keV  respectively, as well as CV, CVI and 
NeIX, NeX K$\alpha$ and K$\beta$ lines are present in these spectra with 
intensities and equivalent widths depending weakly on the ionization state. 
On the other hand, intensities and equivalent widths strongly depend on the 
dynamics of the absorbing  clouds of gas along the line of sight, and on its 
column density. 
For the case under consideration ($\sigma_v = 500$ km s$^{-1}$, $N_H= 
6\times 10^{21}$ cm$^{-2}$) the equivalent widths of K$\alpha$ and 
K$\beta$ He-like and H-like Carbon, Oxygen and Neon are in the range 2-8 eV.
In the medium and higher ionization cases, K$\alpha$ and K$\beta$ lines 
from Mg and Si are also present, along with two complex systems of SXI-SXV, 
and FeXV-FeXX L lines around 0.3 and 1 keV respectively. 
Their equivalent widths in the higher ionization case are of the order of 
1 eV for each line. 
We note that the energies of the strong K$\alpha$ and K$\beta$ NeIX, NeX 
lines, as well as those of a number of FeXVII, FeXVIII lines, are all 
close to each other and to the energy of the OVIII K edge. Caution 
should then be made when fitting low-medium resolution spectra of 
'warm absorbers' with a phenomenological multi-edge model. 
Neglecting the presence of this system of absorption lines (which may 
reach total equivalent width of up to a few tens of eV for $\sigma_v= 
1000$ km s$^{-1}$), and leaving the optical depths of each edge free 
to vary independently in these fits, may result in good fits, from the 
$\chi^2$ point of view, but in overestimating  the OVIII optical depths. 

\subsubsection{Steep X-ray ionizing continuum}

When the ionizing continuum is steep below 2 keV ($\alpha_E\gs 2$) and
flattens above this energy to $\alpha_E\sim0.7-1.5$, as in many NLSy1, the 
resulting ionization structure can be quite different from the ones 
described in the previous section. In fact, in these condition C, O, 
and Ne would be almost fully stripped in the gas, while Iron would still 
be distributed in medium--low ionization states, giving rise to emerging 
spectra with no sharp absorption features below $\sim 1$ keV and a 
broad absorption structure between 1 and 2 keV due to a complex mixture 
of L resonance absorption lines and edges from FeX-FeXX. 

As an example, fig. 4 shows two spectra emerging from an outflowing 
wind photoionized by a typical NLSy1 continuum (see Leighly et al., 
1997), consisting of a black body with kT=0.13 keV, and a power law 
with $\alpha_E = 0.9$. 
{\bf The relative normalization between the soft and the hard components 
is such that the ratio between the bolometric luminosity of the 
black body component and the 2-10 keV luminosity is $(L_{Bol}(BB)/
L_{2-10}(PL)) \simeq 10$ (a mean value in the Leighly et al. sample).}
We assume also in this case $v_{out}=1000$ km s$^{-1}$, $\sigma_v = 500$ 
km s$^{-1}$, and a column of $N_H^{Cold}=3 \times 10^{20}$ cm$^{-2}$ 
neutral gas to mimic a typical Galactic absorption. 
In the upper panel of Fig. 4 we plot the spectrum emerging from gas with 
LogU=-0.5 and Log$N_H$=22. 
Both absorption lines and edges form the H-like and He-like ions of C, 
O and Ne, along with a number of Si L lines around 0.3 keV, are imprinted 
on spectra emerging from photoionized gas with this (or lower) value of U, 
similarly to what is found in the case of flat X-ray ionizing continuum, 
see figure 3.
The lower panel of Fig. 4 illustrates the case of a much higher ionization 
(LogU=1). Here we use a column density an order of magnitude higher than
in the previous case, to highlight the peculiarities of this model: 
no deep absorption edge is visible in this spectrum, despite of the very 
high column of $N_H=10^{23}$  cm$^{-2}$, while a complex system of strong 
absorption lines between 1 and 2 keV is clearly visible, due to almost 60 
L lines of mid-ionized Iron, and K$\alpha$ and K$\beta$ lines of Mg, Si 
and S.
The iron L lines around 1 keV are all very close in energy and their 
equivalent widths span from a few eV, for $\sigma_v = 500$ km s$^{-1}$, up 
to tens of eV for $\sigma_v = 1000$ km s$^{-1}$. The total equivalent
width due to these lines between 1 and 1.5 keV can be as high as of 
few hundreds of eV. 
%
%

Let us now suppose that the ionized gas in NLSy1 has
physical properties similar to the 'warm absorbers' in flat 
X-ray spectrum type 1 AGN, i.e. similar electron density and
similar distance from the source of the ionizing radiation in
units of gravitational radii.
Pounds et al. (1995) and Laor et al. (1997) suggested that
steep X-ray spectrum, narrow line quasars are emitting close to
the Eddington luminosity, at a rate, say, a factor of ten 
higher than flat X-ray spectrum broad line quasars.
This means that the black hole mass of steep X-ray 
quasars is a factor of ten smaller than that of 
flat X-ray quasars with a similar luminosity 
(in agreement with variability studies, Fiore et al. 1998b, and
with detailed modeling of the optical-UV-X-ray SED of 
NLSy1, Puchnarewicz et al 1998, Siemiginowska et al 1998). 
{ \bf The remarkable uniformity of ionization parameters found by
Reynolds (1997) and George et al. (1998), in their samples of Seyfert 1 
galaxies and quasars of luminosity spanning three orders of magnitude 
implies that the product $R^2 n_e$ scales linearly with the
luminosity in these objects. If this correlation holds for NLSy1 too, 
then we can estimate the $R^2 n_e$ factor of a NLSy1 of a given 
luminosity using, for example, the Reynolds estimate but scaling 
the distance between the absorber and the ionizing source by the above 
factor of ten.}
For a NLSy1 of luminosity $10^{43}$ erg s$^{-1}$ we find 
$R^2n_e$$\sim10^{40}$ cm$^{-1}$, using the typical ionization
parameter found by Reynolds (1997). 
Using this value, and the correct ionizing continua, we expect for 
a typical NLSy1 a LogU of 0.8-1.2, and therefore a spectrum
similar to that plotted in the lower panel of Fig. 4.

\section{Data and Simulations}

To compare our models with the available data and to make predictions 
on what we will be able to observe with the next generation of X-ray 
spectrometers (i.e. gratings and calorimeters), we folded our 
photoelectric+resonant absorption model with the responses of the ASCA 
SIS, AXAF ACIS/HETG, and a high resolution, high effective area calorimeter 
as the baseline for the future mission Constellation-X. 
The simulations were performed using model parameters as deduced from the 
ASCA data of the Seyfert 1 galaxy NGC~985, and the NLSy1 IRAS~13224-3809. 
We discuss these two cases in turn.

\subsection{Flat ionizing continuum: The case of NGC~985}

NGC~985 is a Seyfert 1 galaxy known to host an ionized absorber (Brandt et al 
1994). Nicastro et al. (1998b) observed the source with ASCA and 
confirmed the presence of an ionized absorber along the line of sight.
They find that a single-zone photoionization model (accounting for 
photoelectric absorption only,  model {\em a}) is not able to explain 
completely the complex absorption structure seen in the ASCA spectrum
(the continumm has been parameterized as a power law with $\alpha_E=1.3$ 
up to 2.3 keV, while above this energy the spectrum flattens by  
$\Delta\alpha_E=0.6$, the ionized absorber parameters are LogU=$0.93\pm0.09$, 
LogN$_H=21.95\pm0.05$, and z$_{abs} = 0.023^{+0.034}_{-0.029}$, 
which implies a maximum outflowing velocity $\approx 15,000$ km s$^{-1}$). 
Analysis of the ASCA SIS residuals shows a significant deficit of counts 
around $\sim 1 $ keV (figure 5a). 
The equivalent width of this feature (estimated by adding a gaussian 
absorption line to the continuum model) is of -($13\pm 4.0$) eV.

%

We computed the spectrum emerging from gas photoionized by the
NGC~985 optical to X-ray continuum using our photoelectric+resonant 
absorption model.
We extrapolated the soft X-ray power law, as measured from the ASCA 
data, through optical wavelengths. This is consistent with both the 
measured O-UV spectral index $\alpha_{O-UV}=1.4$ (De Vaucouleurs et 
al., 1991) and $\alpha_{OX}$=1.4. 
We adopted for LogU and Log$N_H$ the best fit values given above 
and an outflowing velocity of 1000 km s$^{-1}$, which is the mean 
measured value for the known O-UV/X-ray ionized absorbers (Mathur et al., 
1994, 1995, 1997, 1998), and is consistent with the estimate of the 
'warm absorber' redshift in this source. We used a dispersion velocity of 1000 
km s$^{-1}$. The  model is plotted in figure 5c. It shows
strong resonance absorption lines between 0.85 and 1.2 keV due to 
the redshifted K$\alpha$ and K$\beta$ NeIX and K$\alpha$ NeX lines.
The blend of these three lines has a total equivalent with of about
20 eV, similar to the equivalent width of the line feature in figure 5a.
We folded this model with the ASCA--SIS response, and simulated a 
40 ksec observation of NGC~985. We fitted the resulting spectrum
with model {\em a}). The ratio between the simulation and best fit model
{\em a}) is shown in figure 5b: it resembles strikingly the residuals to the 
`real' spectrum in figure 5a. While this is certainly not 
a self consistent, quantitative and unique description of the 
spectrum of NGC~985, it indicates that resonant absoprtion may well be 
cause of the feature observed at about 1 keV.

The spectrum in figure 5c also shows strong absorption K$\alpha$ 
and K$\beta$ lines from OVII and OVIII at 0.57 and 0.65  keV. 
There is no evidence for an absorption feature (in addition to 
photoelectric absorption) at these energies in  the ASCA spectrum of 
NGC~985 (Fig. 5a). 
Contribution from emission lines due to recombination could be important 
at these energy, and so they could help in reducing the depth of the 
absorption lines. 
Line emission strongly depends on the covering factor and therefore
an accurate estimate of line emission intensity can provide information
on the geometry and distribution of the ionized gas.
However, the above energy is close to the lower threshold of the SIS 
instrument, where the calibration is most uncertain and problems
due to the Residual Dark Distribution (RDD) are large, and where
the CCD energy resolution is of 5-6 only. 
For a quantitative study of the OVII and OVIII emission/absorption
lines we should await instruments with better energy resolution and 
coverage.

\bigskip
As a next step we folded the photoelectric+resonant absorption 
model of figure 5c with the responses of AXAF ACIS/HETG and 
the baseline Constellation-X calorimeter, which has a factor of
2 worse energy resolution but a factor of 100 higher effective area 
(at 1 keV). 
The simulations are shown in figures 6a,b,c and 7a,b,c for three different
values of the dispersion velocity: $\sigma_v=100$, 500 and 1000 km s$^{-1}$
for panels a,b and c respectively.
The first order AXAF ACIS/MEG (``Medium Energy Grating'') simulations 
(figure 6) were performed using the MARX simulator (Wise et al., 1997, 
``MARX User Guide'', vs 1.0)
\footnote{http://asc.harvard.edu/prop\_port.html\#marker\_marx}
). The exposure time is of 80 ksec. 
To simulate the calorimeter spectrum (figure 7), we wrote a simple 
program to fold our models with the response matrix 
\footnote{ftp://legacy.gsfc.nasa.gov/htxs/calor.rsp}
and add statistical noise.
The exposure time in the Constellation-X calorimeter simulations is of 
10 ksec. 

Resonance absorption lines of width $\gs 200$ km s$^{-1}$ are detected 
in the AXAF ACIS/HETG first order spectrum. 
All the strongest lines in figure 5c are resolved in the spectra of 
figure 6b,c. AXAF HETG spectra of Seyfert 1 galaxies 
will then establish whether resonant absorption is important in the
ionizing gas and then provide estimates of the gas dispersion
velocity and bulk motion. This will allow to solve unambiguously the 
open and controversial issue on the identification between the X-ray 
and the UV ionized absorbers.  
However, the signal to noise will be poor in most cases:
the OVIIK$\alpha$/K$\beta$ and NeIX K$\alpha$/K$\beta$ line ratios, 
which can be used for plasma diagnostic, cannot be estimated with
uncertainty $\ls 100\%$; line widths and profiles, which gives information 
on the geometrical and dynamical configuration of the absorber can be 
roughly measured but not studied in detail.  
The HETG can obtain spectra of comparable quality for about one hundred AGN.

The large effective area of the Constellation-X calorimeter 
along with a good energy resolution will permit to obtain high signal 
to noise spectra, of quality similar to optical spectra of AGNs, 
of sources of the brightness of NGC~985 . This will allow to apply plasma 
diagnostics, measure accurately both the chemical composition and 
ionic distribution of the gas, discriminate between collisional ionization 
and photoionization, and study the time-evolving behaviour of the relevant 
physical quantities (temperature, relative ionic abundances, etc.) which 
will allow in turn to constrain important physical and geometrical 
parameters (electron density, and radial distance: Nicastro et al., 1998a). 
The Constellation-X calorimeter could provide moderate signal to noise 
spectra, similar to those in figure 6, of sources down to a flux limit of 
a few $10^{-13}$ erg cm$^{-2}$ s$^{-1}$, which increases the sample of
possible targets to a few tens of thousands sources. 
AGNs with luminosity in the range $10^{43-44}$ 
erg s$^{-1}$ can be studied up to z=0.3-0.5, while 
more luminous ($10^{46-47}$ erg s$^{-1}$)  
quasars will be available up to z$\sim3$ (the limit is given by the
bandwidth of the Constellation calorimeter more than by its
sensitivity, since features at 0.5-1 keV are redshifted
to 0.12-0.25 keV for z=3).
This will permit to study the evolution of physical state, chemical 
composition, and geometry of the absorbing gas with redshift and 
luminosity. Statistical studies on the distribution of the absorber
density, ionization state, chemical composition and distance from
the ionizing source will be possible, which can give information 
on the absorber origin.

%
%
%
%

\subsection{Steep Ionizing Continuum: the Case of IRAS~13224-3809}

We choose to test our model on the NLSy1 galaxy IRAS~13224-3809, 
which showed in an ASCA observation a deep absorption feature in the 
interval 1-2 keV, interpreted by Leighly et al. (1997) in terms of 
resonant absorption from gas outflowing with v/c$\sim0.5$

{\bf Figure 8a shows the ratio between the coadded ASCA SIS0 and SIS1 data and 
the best fitting model composed by a black body plus a power law, with 
best fit parameters given by kT=0.134 keV, $\alpha_E=1.16$ and $(L_{Bol}(BB)/
L_{2-10}(PL)) = 9.8$ (fully consistent with both the Leighly et al, 1997, and 
the Hayashida et al., 1997, estimates).} 
A deep and broad deficit of counts is evident between 0.9 and 2 keV (also 
see figure 1 in Leighly et al. 1997). The absorption feature 
equivalent width (estimated by adding a gaussian absorption line
to the continuum model) is large: EW=-(180$\pm30$) eV. 
{\bf We note that, even if the deficit of counts occurs approximately 
where the black body and the power law components match each other, 
we were not able to model the absorption--like feature without invoking 
a third component.} 

%
%

We computed the spectrum emerging from gas photoionized by the 
IRAS~13224-3809 UV to X-ray continuum for a number of values of LogU 
(from -0.5 to 1.5), LogN$_H$ (from $10^{22}$ to $10^{24}$ cm$^{-2}$) 
and $\sigma_v$ (from 100 to 1000 km s$^{-1}$). 
We extrapolated down to 13.6 eV and up to to {\bf 100 keV} the best fit 
continuum to the ASCA data of IRAS~13224-3809. This is in good 
agreement with the optical UV to X-ray measures of this source which 
suggest that the thermal disk component in this object could be present 
at higher energies than usually found in soft X-ray flat, broad line 
type 1 AGN (Mas-Hesse et al., 1994). 
{\bf We adopted, for the hard X-ray power law, the best fit spectral index 
$\alpha_E = 1.16$. This value is in excellent agreement with the mean 
estimate of Hayashida et al. (1997). However, as Brandt, Mathur 
\& Elvis (1997) pointed out, the spectral index of the hard power law 
of IRAS~13224-3809 is highly variable, ranging between 0.7 and 1.3. 
The value that we used in our model represents therefore only a mean value. 
We stress however that the exact shape of the hard power law does not 
affect strongly the distribution of the relative ionic abundances in the 
gas for elements lighter than Neon, which is instead mainly driven by the 
presence of a strong soft excess below $\sim 2$ keV. 
Our results are therefore rather insensitive to the exact slope of the 
hard power law.} 
As in the previous case we adopted an outflowing velocity of
1000 km s$^{-1}$. We folded these models with the ASCA-SIS response 
and then fitted them with a black body plus power law model. As an 
example figure 8b shows the ratio between an 80 ksec simulation obtained 
using log$N_H$=23.5, LogU=1 and outflowing and dispersion velocities of 
1000 km s$^{-1}$ (model shown in figure 8c), and the best fitting black 
body plus power law model. 
{\bf We note that all the relevant absorption lines lie in a spectral 
region in which the intrinsic continuum is still dominated by the 
soft black body component (Fig 8c, but see also Fig. 4).} 
The residuals resemble those in figure 8a. The large deficit of counts
between 0.9 and 2 keV is due to the presence of a large number
of iron L lines. The coadded equivalent width of these lines is about
130 eV, similar to the equivalent width of the absorption feature in 
figure 8a.

\bigskip
As for the case of NGC~985, we plot in figure 9a,b,c and 10a,b,c 
three simulations obtained convolving the spectrum in figure 8c with the 
AXAF/MEG and Constellation-X calorimeter responses. 
The simulations are shown for three values of the dispersion velocity, 
$\sigma_v=100$, 500 and 1000 km s$^{-1}$ (figure a, b and c respectively). 
Integration time was 100 ksec for the first order AXAF/MEG spectrum and 
20 ksec for the Constellation-X calorimeter spectrum.

Many single Fe L lines are well resolved even in spectra with dispersion
velocity of 1000 km s$^{-1}$. AXAF/HETG spectra of low redshift NLSy1 with 
luminosity L$=10^{43}-10^{44}$ erg s$^{-1}$ will permit to unambiguously 
establish whether an ionized absorber is indeed responsible for the 1 keV 
deficit of counts seen in many moderate resolution X-ray spectra of these 
sources. 
The low signal to noise ratio of AXAF/HETG spectra, however, will not 
allow to apply plasma diagnostics to infer the physical properties of 
the absorbers. 
Nevertheless the outflowing and dispersion velocities of the gas will 
be accurately derived, as well as the relative Ne to O, and Fe to Si and 
Mg abundances, which will put important constraints on the geometry and 
the chemical composition in the AGN environment. 
The high effective area of the Constellation-X calorimeter will allow to 
greatly increase the sample of narrow-line type 1 AGN observed at high 
resolution. The complex Iron L edges and lines absorption system could 
be detected in high luminosity narrow line quasars up to redshift $\ls2$,  
or even higher, depending on the actual response below the carbon edge 
(0.283 keV), while Mg and Si K$\alpha$ and K$\beta$ resonance 
absorption lines could be detected up to z$\sim 3$. 
The good energy resolution (3 eV), will permit to separate the strongest 
(oscillator strength greater than 0.05) single resonance absorption lines 
of the 1 keV Iron L edge+line complex, so allowing for a clear diagnostic 
of the mechanism responsible for the ionization of the gas. 
In particular, an accurate measure of the relative abundances of the 
very stable FeXVII Li-like ion could allow to clearly distinguish between 
gas in collisional or photoionization equilibrium (Nicastro et al. 1998a).
Time evolving ionization can also be investigated, so giving constraints 
on the electron density and the radial distance of the absorbers (Nicastro 
et al. 1998a). 

%
%
%
%

\section{Conclusions}

We have calculated spectra emerging from ionized gas taking into
account both photoelectric and resonant absorption.
We find that the emerging spectra are strongly modified by resonant
absorption if the dispersion velocity in the gas is $\gs 100$ km s$^{-1}$.
This or even higher values are not irrealistic, since many 
broad line AGN show absorption lines in the UV blueshifted by 
200-2000 km s$^{-1}$, and broadened by similar dispersion velocities.  

In particular we find:

\begin{enumerate}

\item The distribution of relative ionic abundances in photoionized 
gas dramatically depends on the exact shape of the ionizing continuum. 

\item Strong K$\alpha$ and K$\beta$ resonance absorption lines from He-like 
and H-like ions of C, O, and Ne, along with deep CVI, OVII-OVIII, NeIX-X K 
edges, and FeXV-XVII L edges, are expected to be present in the 0.1-3 keV 
spectra of flat X-ray spectrum quasars emerging from photoionized gas along 
the line of sight. When fitting low and medium resolution X-ray data with 
phenomenological multi-edge models, these resonance absorption lines may 
produce the effect of overestimating the optical depth of the OVIII K edge.  

\item Conversely, no strong absorption edge is expected to be imprinted 
on spectra emerging from gas illuminated by steep ionizing continua, with 
physical and geometrical pro\-per\-ties similar to those found in broad line 
type 1 AGN. However, a large number of strong Fe L and Mg, Si, and S 
K$\alpha$ and K$\beta$ resonance absorption lines (almost 70) 
are predicted between 1 and 2 keV. Fitting low and medium energy 
resolution data with continuum models may then result in deep smooth 
negative residuals at these energies.
We then suggest that the $\sim 1$ keV absorption feature observed 
by ASCA in several Narrow Line Seyfert 1 galaxies can be explained
in terms of resonance absorption lines, without requiring relativistic 
outflowing velocities of the gas.

\item The equivalent width of the predicted resonant absorption lines
strongly depends on the column density of the absorbing gas and 
the its dispersion velocity.
Assuming the range of N$_H$ typically observed and dispersion velocities of 
0--1000 km s$^{-1}$, we estimate equivalent widths spanning from a few 
eV to a few hundreds of eV. 

\end{enumerate}

\bigskip
F.F. acknowledges support from NASA grant NAG5-2476 and 
NAG5-3039. 


\newpage
%
\figcaption{Growth curves of the K$\alpha$ and 
K$\beta$ lines of H-like and He-like C, O, Ne and Fe, for $\sigma_v = 0$. 
The ionization state of the absorbing gas is that of a typical 
'warm absorber' (LogU=1) for C, O, and Ne, and a much higher ionization 
case (LogU=3) for FeXXV and FeXXVI.}
%

%
\figcaption{As Fig. 1, but for $\sigma_v = 1000$ km s$^{-1}$.}
%

%
\figcaption{0.1--5 keV spectra emerging from an 
outflowing wind photoionized by a typical flat X-ray, broad line type 1 AGN 
continuum, for LogU = 0.5, 1, 1.5, $v_{out} = 1000$ km s$^{-1}$, 
$\sigma_v=500$ km s$^{-1}$ and N$_H = 6 \times 10^{21}$ cm$^{-2}$. 
The low energy turn over is due to absorption from a column of neutral gas, 
to mimic typical Galactic absorption.}
%

%
\figcaption{Spectra emerging from an outflowing 
wind photoionized by a typical NLSy1 continuum, and with $v_{out} = 
1000$ km s$^{-1}$ and $\sigma_v=500$ km s$^{-1}$. 
The ionizing continuum used consists of a black body with kT=0.13 keV, 
and a power law with $\alpha_E= 0.9$. The ionization parameter and
gas column density are LogU=-0.5, Log$N_H$=22, and LogU=1, Log$N_H=23$ 
in the upper and lower panels respectively.}
%

\figcaption{Ratio between the coadded ASCA SIS0 and SIS1 data of NGC985 
and the best fitting model {\em a} (a). 
Ratio between a 40 ksec ASCA SIS simulation, obtained using the model 
plotted in the lower panel (see the text for details), and model {\em a} (b).
The spectrum emerging from gas photoionized by the NGC~985 SED, 
when accounting for both photoelectric and resonant absorption 
(c). We adopted  outflowing and dispersion velocities of 1000 km s$^{-1}$.}
%

%
\figcaption{Simulations of 80 ksec 0.4--1.2 keV first order 
AXAF ACIS/MEG spectra of NGC 985.  A dispersion velocity of $\sigma_v=100$, 
500 and 1000 km s$^{-1}$ has been used in figure a), b) and c) respectively.}
%
%
\figcaption{Simulations of 10 ksec spectra of NGC~985  
taken with the Constellation-X  calorimeter. A dispersion velocity of 
$\sigma_v=100$, 500 and 1000 km s$^{-1}$ has been used in figure a), b
and c) respectively.
The cut-off at low energy is partly due to the inclusion at low energy of
neutral absorption from the Galactic column along the line of sight 
($N_H=3\times10^{20}$ cm$^{-2}$).}
%

%
\figcaption{Ratio between the coadded ASCA SIS0 and SIS1 
data of IRAS~13224-3809 and the best fitting black body  plus power law 
model (a). 
Ratio between a 80 ksec ASCA SIS simulation, obtained using the model 
plotted in the lower panel (see the text for details) and the best
fitting black body plus power law model (b).
The spectrum emerging from gas photoionized by the IRAS~13224-3809 SED, 
when accounting for both photoelectric and resonant absorption 
(c). We adopted outflowing and dispersion velocities of 1000 km s$^{-1}$.}
%

%
\figcaption{Simulations of 100 ksec 0.4-2 keV first order 
AXAF ACIS/MEG spectra of IRAS~13224-3809.  
A dispersion velocity of $\sigma_v=100$, 500 and 1000 km s$^{-1}$ has 
been used in figure a), b) and c) respectively.}
%
%
\figcaption{Simulations of 20 ksec 0.3--2 keV spectra of IRAS~13224-2809 
taken with the Constellation-X calorimeter. A dispersion velocity of 
$\sigma_v=100$, 500 and 1000 km s$^{-1}$ has been used in figure a), b) 
and c) respectively.
The cut-off at low energy is partly due to the inclusion of neutral 
absorption from the Galactic column along the line of sight ($N_H=4.8
\times10^{20}$ cm$^{-2}$).}
%

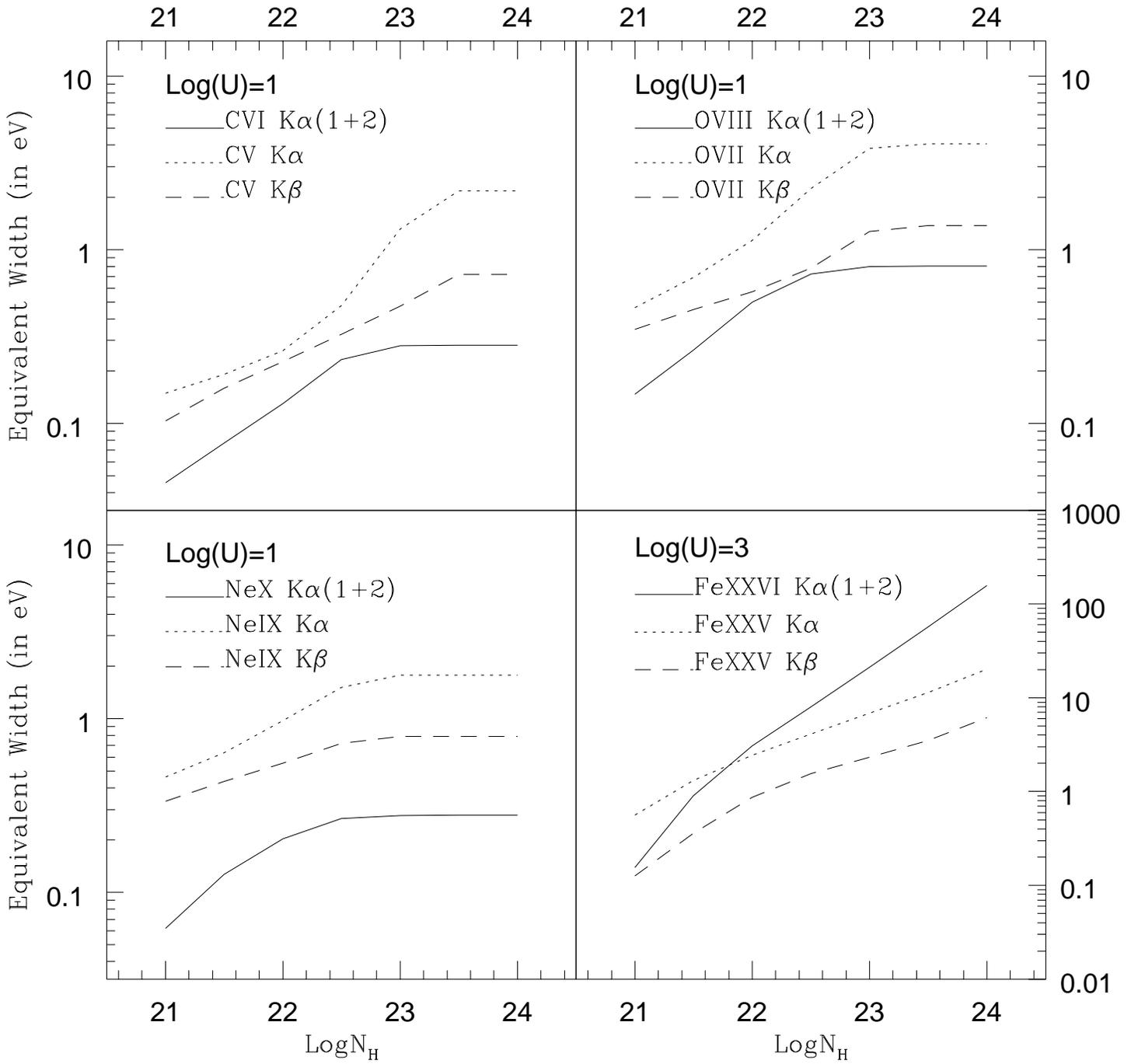
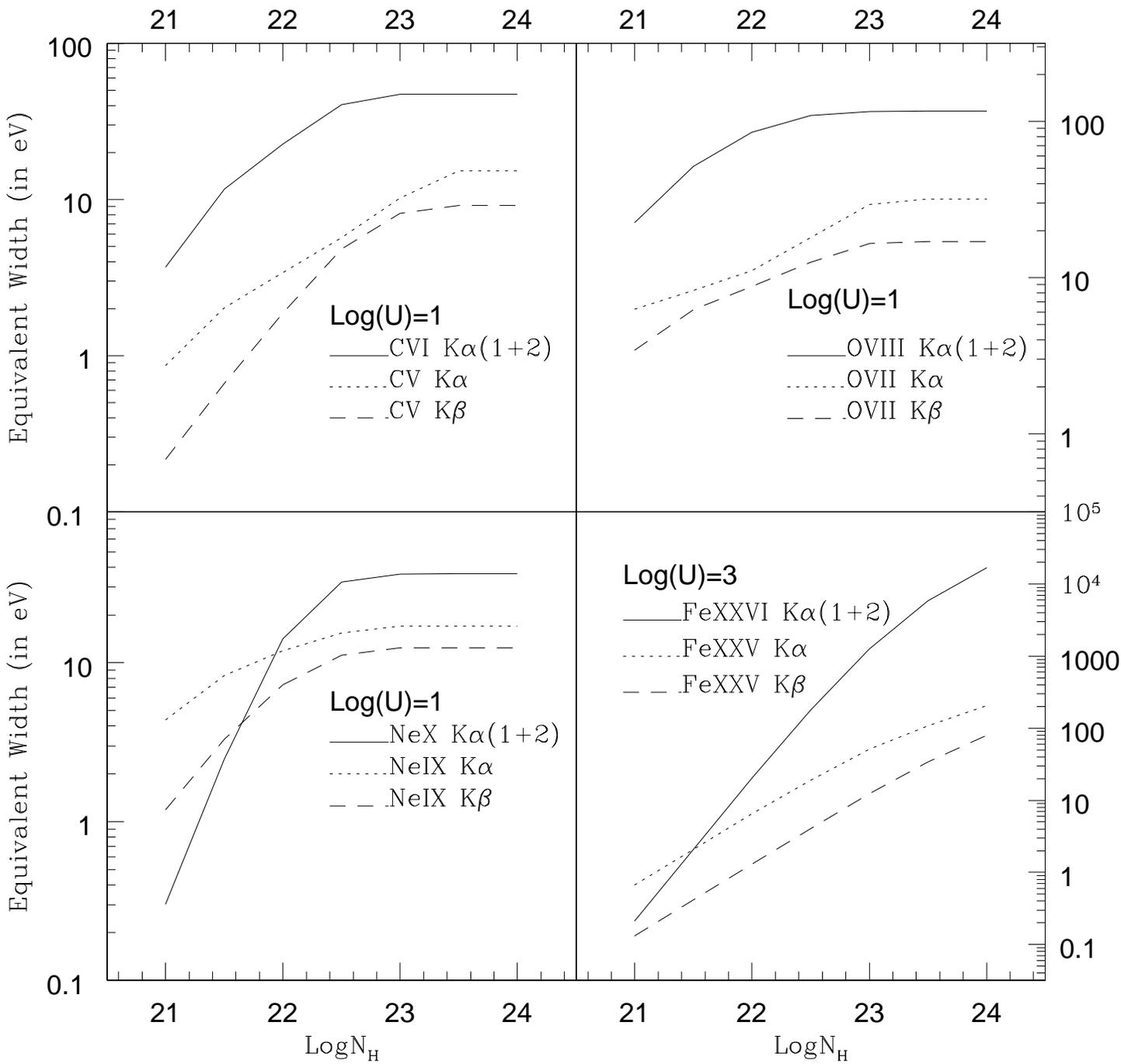
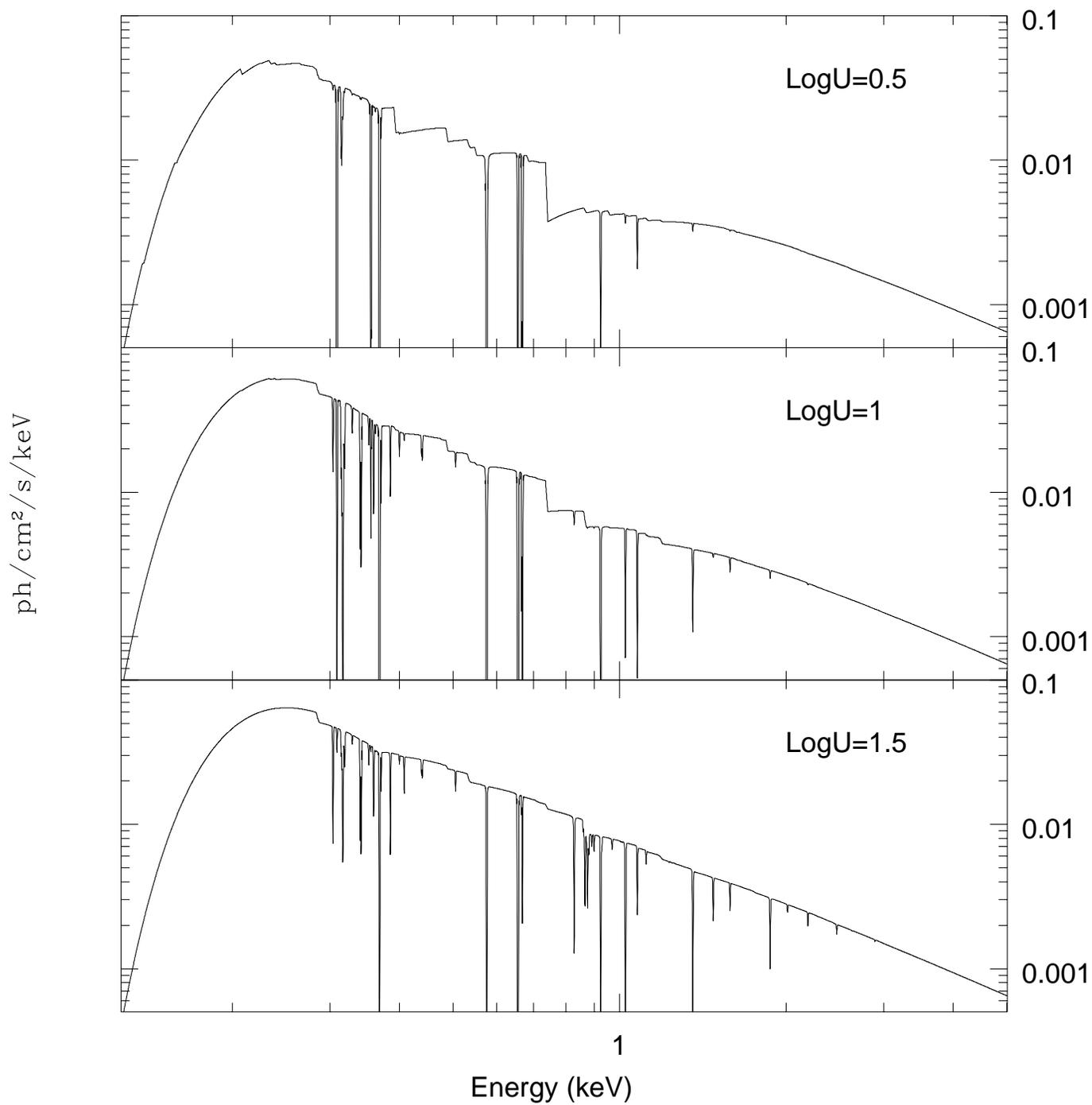
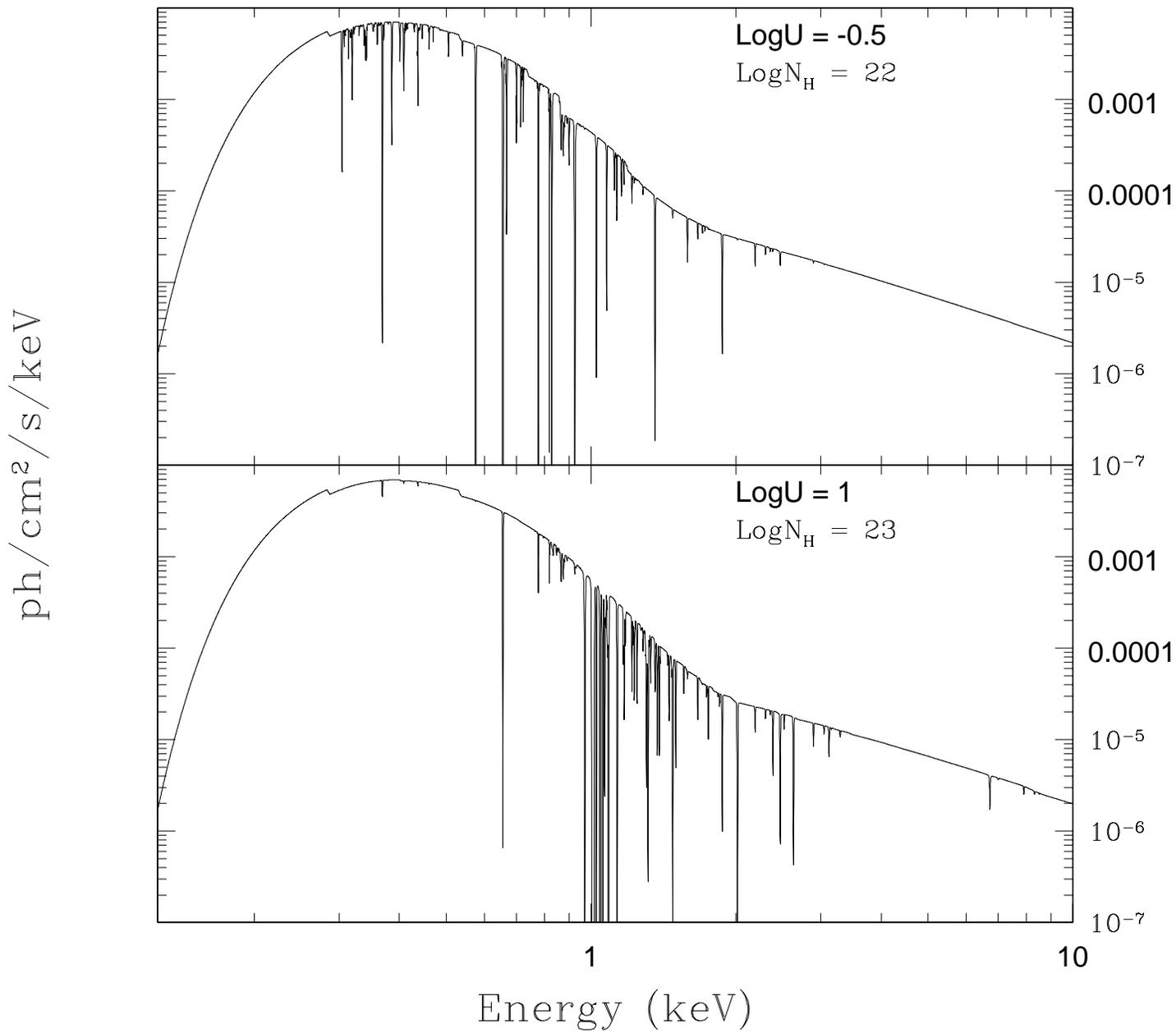
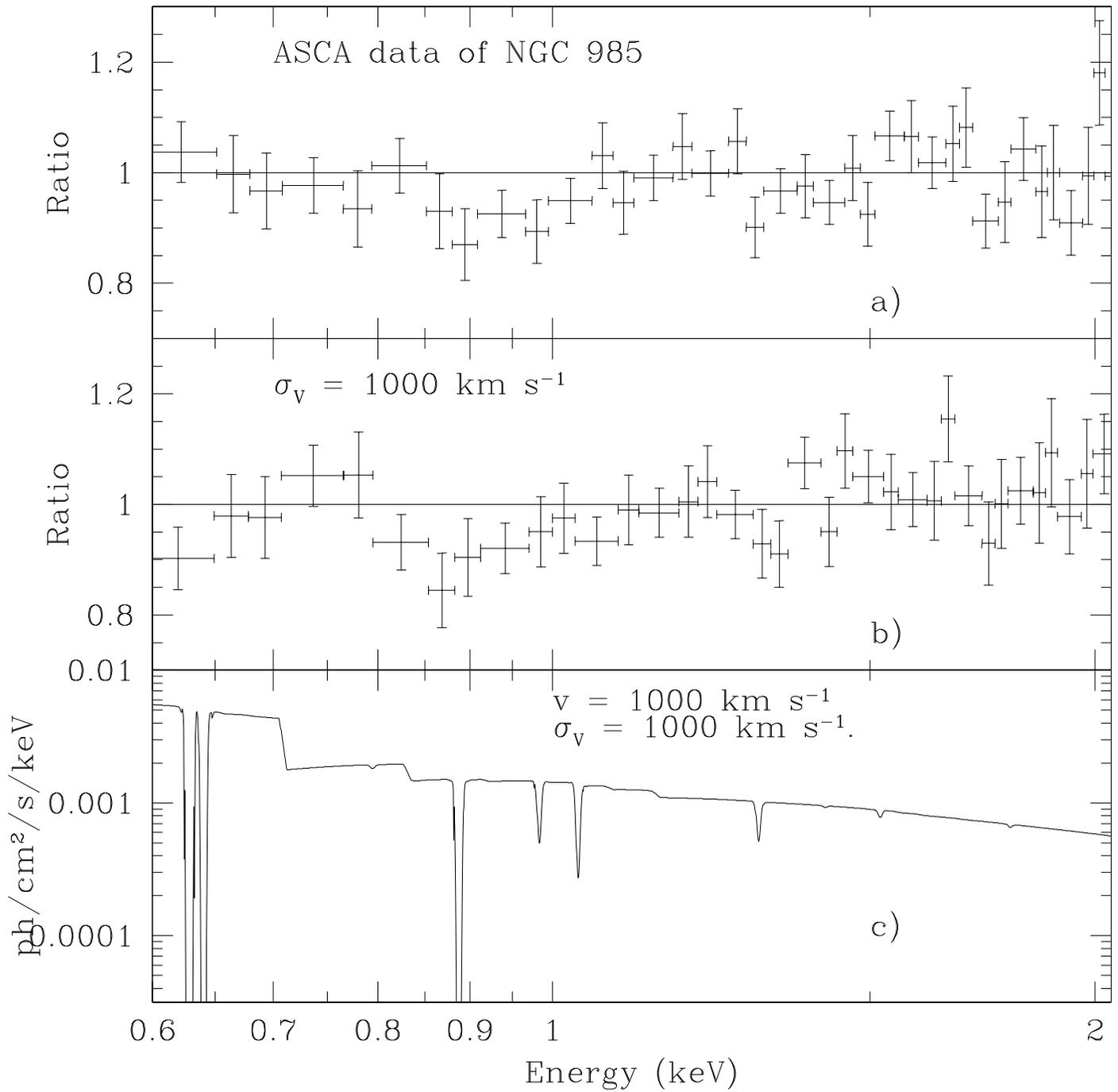
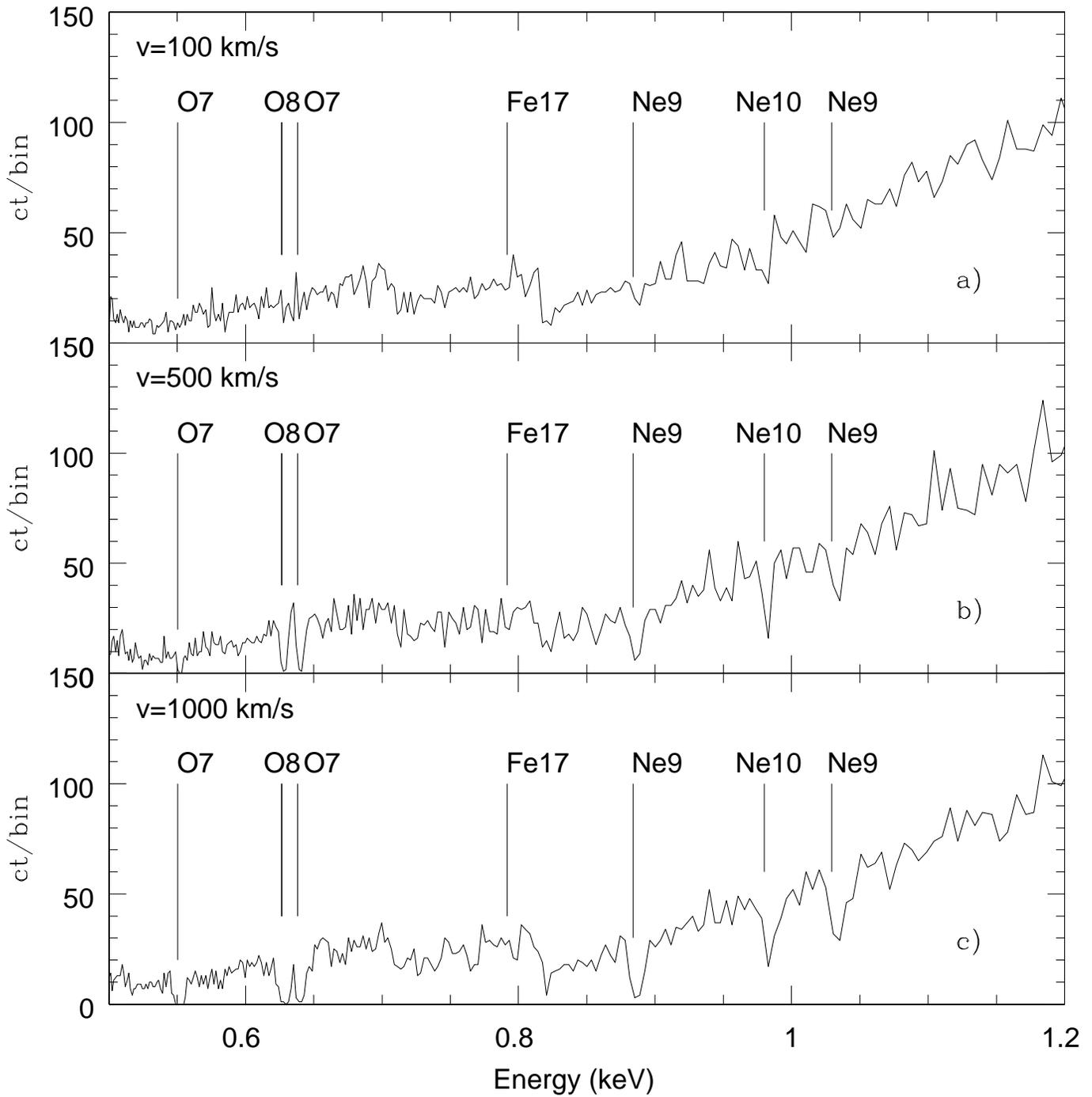
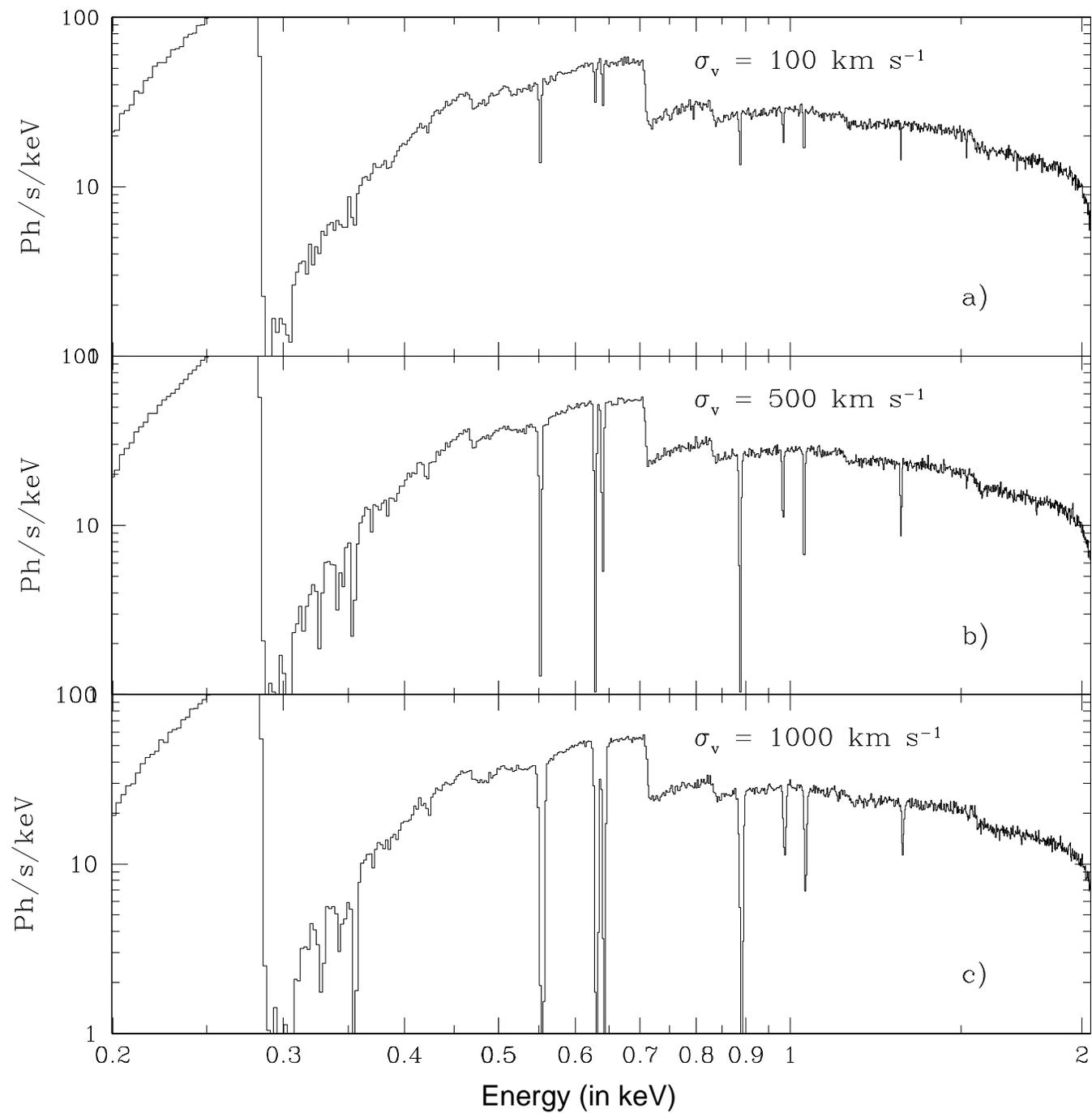
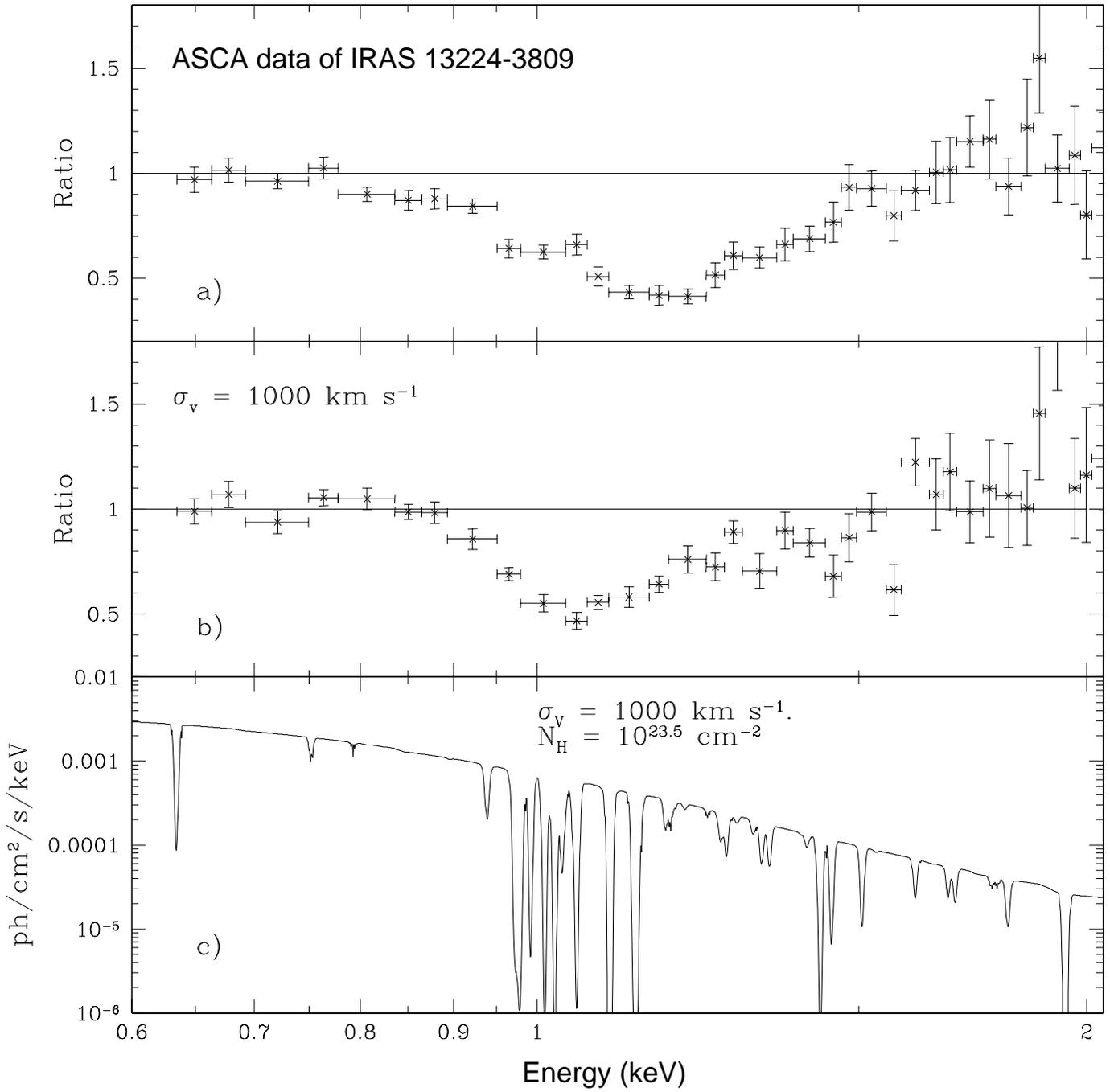
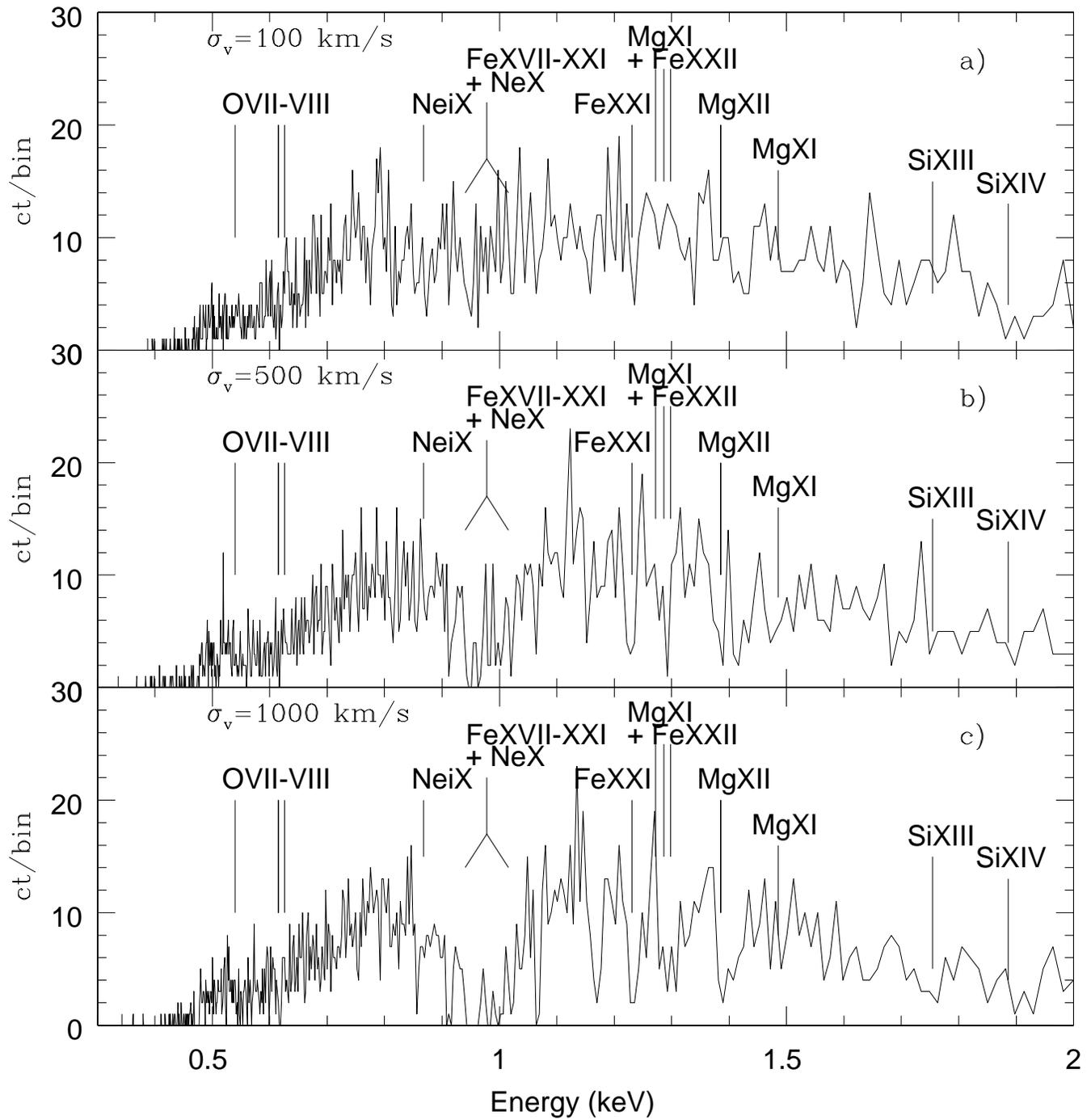
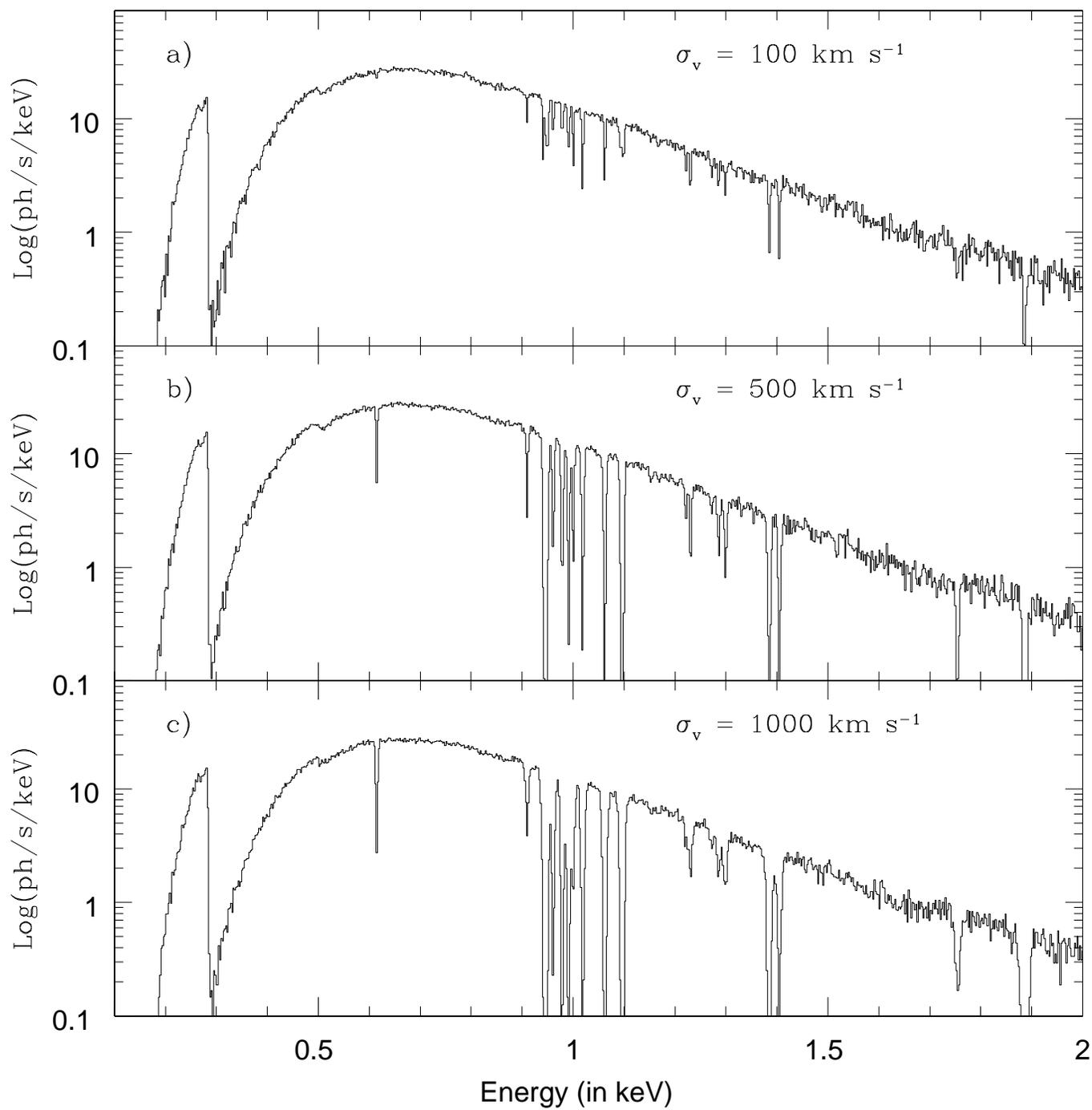
\end{document}